# Current Distribution in a Two-Dimensional Electron Gas Exposed to a Perpendicular Non-Homogeneous Magnetic Field of a 'Chess' Configuration

Samvel M. Badalyan[a)] and Francois M. Peeters[b)]

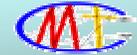 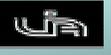

[a)] *Department of Radiophysics, Yerevan State University, 375025 Yerevan, Armenia*
[b)] *Department of Physics, University of Antwerp (UIA), 2610 Antwerpen, Belgium*

## Motivation

Effects induced by the geometry and by non-homogeneous magnetic fields in low dimensional semiconductor systems are of current interest [1,2]. Energy dissipation processes and transport properties of these systems are mainly determined by the spatial distribution of the electric current and the potential. These distributions are essentially altered by the sample geometry (the sample shape and size, the electrode configuration) and by the non-homogeneity of the applied magnetic field.

We calculate the Joule heat generation rate in a finite 2DEG exposed to a normal non-homogeneous magnetic field $B(x,y)$ of a 'chess' configuration.

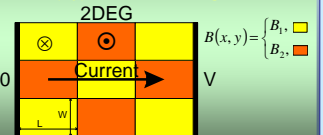

$$B(x,y) = \begin{cases} B_1 \\ B_2 \end{cases}$$

The number of 'chess' fields is arbitrary.

## Analytical Results

For the anti-symmetric system with $B_1 = -B_2 = B$ we obtain the following exact solution for the electric field distribution

$$-iE^*(z) = -iE_x - E_y = \left[\frac{\left(1-w\right)\frac{1}{\sqrt{m}} + w}{\left(1+w\right)\frac{1}{\sqrt{m}} - w}\right]^{\frac{|\delta|}{\pi}}, \quad w = sn\left(\frac{z}{c}, m\right)$$

where $z = x + iy$, $\delta$ is the Hall angle, $sn$ the Jacobi elliptic function, $K(m)$ and $K'(m)$ are the complete elliptic integrals of the first kind, the $m$ index and the $c$ constant are determined by the sample aspect ratio from $W = 2cK(m)$ and $L = cK'(m)$.

## Electric Field

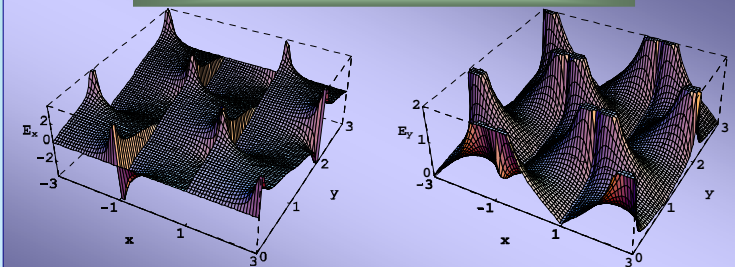

**Fig. 1.** The spatial distribution of the electric field components $E_x(x,y)$ and $E_y(x,y)$ for the non-homogeneous magnetic field distribution corresponding to the 3⊗3 'chess' configuration.

## Theoretical Approach

Assume that an electric potential is applied at the ends of the 2DEG, which is in the (x,y)-plane, and a constant current flows in the y-direction.
In the steady state the electric field and the current density are determined by the Maxwell and the continuity equations supplemented by Ohm's law

$$\text{rot}\, E(x,y) = 0, \quad \text{div}\, J(x,y) = 0, \quad J_\alpha(x,y) = \sigma_{\alpha\beta}(x,y) E_\beta(x,y)$$

where $\sigma_{\alpha\beta}(x,y)$ is the spatial dependent magneto-conductivity. This set of equations has to be solved with the boundary condition

$$\left.\begin{array}{l} E_{\text{tangential}}(x,y) \\ J_{\text{normal}}(x,y) \end{array}\right\} \text{are continuous.}$$

The electric potential of this system *does not* satisfy the Laplace equation in the whole 2DEG region. The line charges are accumulated at the magnetic interfaces and the potential satisfies the Poisson equation with a non-trivial right side part, which is determined by the field itself.

- In each 'chess' field, however, the magnetic field is uniform and the potential satisfies the 2D Laplace equation.
- We obtain an electric field distribution in each 'chess' field using a conformal mapping method.
- Our approach is different in the second step from that used in Ref. [3].
- We introduce additional unknown angles, which the electric field makes with respect to the magnetic interfaces at each 'chess' field boundary.
- We find a solution of the electrostatic problem in terms of the Jacobi elliptic functions.
- We match these solutions at the magnetic interfaces and eliminate the above angle.

## Joule Heat

## Line Charges

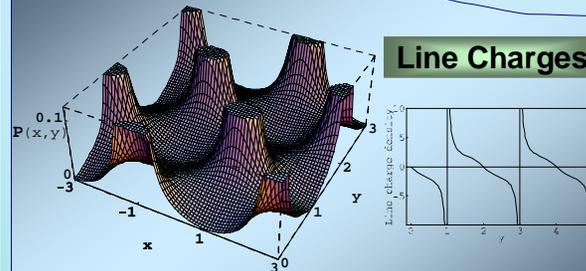

**Fig. 3.** Joule heat generation rate $P(x,y) = E(x,y) \cdot J(x,y)$ (in arbitrary units) corresponding to the situation of Fig. 1 (the left figure). Line charge density accumulated at the magnetic interfaces along the y-direction (the right figure).

## Current density

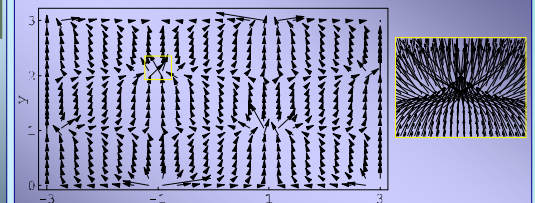

**Fig. 2.** Contour plot of the current density distribution corresponding to the situation in Fig. 1. A singular point at the current injection and removal corner (the right figure). The current density has the following analytical behavior: $J(z) \propto (z - z_{corner})^{-2|\delta|/\pi}$.

## Summary

We have calculated analytically the spatial distribution of the electric field and current density, and obtained the Joule heat generation rate in a two-dimensional electron gas (2DEG) subjected to a non-homogeneous magnetic field of a 'chess' configuration. The generation of the Joule heat from the 2DEG is mainly concentrated near the singular corners of each 'chess' field and tends to zero in other corners.

## Acknowledgments

This work was partially supported by the Flemish Science Foundation (FWO-VI), the Inter-University Attraction Poles research program (IUAP-IV), the Concerted Action Program (GOA), and the Inter-University Microelectronics Center.
S.M.B. acknowledges support by the Vatche and Tamar Manoukian Benevolent Association (VTMBA) to participate the 'Phonons 2001' conference, August 12-17, 2001, Hanover, USA.